\pdfoutput=1

\documentclass{PoS}

\usepackage{amsmath}
\usepackage{eucal}
\usepackage{mathrsfs}          

\newcommand{\eg}{e.g.~}
\newcommand{\ie}{i.e.~}

\newcommand{\Fig}[1]{Fig.~\ref{#1}}

\newcommand{\Op}{\CMcal{O}}		

\newcommand{\ZZ}{\mathbb{Z}}

\newcommand{\beq}{\begin{equation}}
\newcommand{\eeq}{\end{equation}}

\newcommand{\SUd}{SU(2)_{\text{L}}}

\newcommand{\FERMI}{{\sf Fermi}}

\newcommand{\HESS}{{\sf H.E.S.S.}}

\newcommand{\LUX}{{\sf LUX}}

\newcommand{\X}{\mathcal{X}}
\newcommand{\mDM}{M}

\def\hhref#1{\href{http://arxiv.org/abs/#1}{#1}} 
\def\doi#1#2{\href{https://dx.doi.org/#1}{#2}} 

\title{Generalizing Minimal Dark Matter: Millicharge or Decay\footnote{Based on Ref.~\cite{DelNobile:2015bqo}.}}

\ShortTitle{Generalizing Minimal Dark Matter: Millicharge or Decay}

\author{\speaker{Eugenio Del Nobile}
      \\
      Dipartimento di Fisica e Astronomia ``Galileo Galilei'', Universit\`a di Padova, \\ and INFN Sezione di Padova, Via Marzolo 8, 35131 Padova, Italy
      \\
      E-mail: \email{delnobile@cp3-origins.net}}

\abstract{
The Minimal Dark Matter framework classifies viable Dark Matter (DM) candidates that are obtained by simply augmenting the Standard Model of particle interactions with a new multiplet, without adding new ad hoc symmetries to make the DM stable. The model has no free parameters and is therefore extremely predictive; moreover, recent studies singled out a Majorana $\SUd$ quintuplet as the only viable candidate. The model can be constrained by both direct and indirect DM searches, with present time gamma-ray line searches in the Galactic Center being particularly sensitive. It is therefore timely to critically review this paradigm and point out possible generalizations. We propose and explore two distinct directions. One is to lower the cutoff of the model, which was originally fixed at the Planck scale, to allow for decays of the DM quintuplet. We analyze the decay spectrum of this candidate in detail and show that gamma-ray data constrain the cutoff to lie above the GUT scale. Another possibility is to abandon the assumption of DM electric neutrality in favor of absolutely stable, millicharged DM candidates. We explicitly study a few examples, and find that a Dirac $\SUd$ triplet is the candidate least constrained by indirect searches.
}

\FullConference{The European Physical Society Conference on High Energy Physics\\
		 5-12 July, 2017
		 \\
		 Venice, Italy}

\begin{document}

\section{Introduction}
Among the few things we know about Dark Matter (DM) is that it must be stable on cosmological timescales. In terms of particle physics, stability means symmetry: there must be a symmetry, exact or approximate, responsible for the absence or suppression of Lagrangian operators which may cause the DM to decay. The simplest example is a global $\ZZ_2$ or $U(1)$ symmetry if the DM field is or is not self-conjugated, respectively.

A common way of enforcing DM stability in model building is to impose such a symmetry by hand, postponing the issue of justifying its ultraviolet origin. A different, elegant way to ensure stability is instead exploiting accidental symmetries, the same mechanism that makes the proton stable in the Standard Model (SM). Accidental symmetries are global symmetries appearing in a renormalizable theory as a consequence of its specific matter content, without being imposed a priori.

This is the main idea behind the Minimal Dark Matter (MDM) setup, first presented in Ref.~\cite{Cirelli:2005uq} (see also Refs.~\cite{Cirelli:2007xd, Cirelli:2009uv}). There, the SM is augmented with a new generic multiplet $\X$ with generic quantum numbers under the SM gauge group, without introducing new symmetries. The multiplet mass, the model's only free parameter, is fixed by requiring the DM to be a thermal relic. In listing all possible scalar and spin-$\frac{1}{2}$ candidates, one must take into account the following facts~\cite{Cirelli:2005uq}:
\begin{itemize}
\item Colored thermal relics are very constrained.
\item DM candidates with tree-level interactions with the photon and/or the $Z$ boson are ruled out by direct detection experiments. Therefore only odd-dimensional representations of $\SUd$ are viable choices (see however Refs.~\cite{Hisano:2014kua, Nagata:2014aoa}).
\item Multiplets for which Yukawa couplings with SM fields exist, making the DM unstable, are to be discarded. Also dimension-$5$ operators, in an Effective Field Theory (EFT) approach, make the DM decay too quickly even for a cutoff at the Planck scale.
\item Matter charged under representations of $\SUd$ with larger and larger dimension make the $\SUd$ coupling constant run faster and faster. For large enough representations, a Landau pole makes it necessary to modify the low-energy theory, effectively lowering the EFT cutoff thus making higher-order EFT operators potentially dangerous in inducing fast DM decay.
\end{itemize}
All in all, enforcing electric neutrality for the DM, only one viable candidate is singled out: a fermion $\SUd$ quintuplet with zero hypercharge and mass $\mDM \approx 9$ TeV.\footnote{A more recent analysis including the effects of bound-state formation finds $\mDM \approx 11.5$ TeV~\cite{Mitridate:2017izz}. We neglect these effects in the following.} A DM candidate previously believed to be viable, part of a scalar eptaplet, actually decays too quickly due to a previously overlooked dimension-$5$ operator trilinear in $\X$~\cite{DelNobile:2015bqo, DiLuzio:2015oha}.

At present, direct DM detection experiments are not sensitive to the quintuplet MDM model due to the large radiative mass splitting among multiplet components and the loop-suppressed elastic DM-nucleus scattering cross section (see \eg Ref.~\cite{DelNobile:2013sia}). However, experiments like the \FERMI~{\sf LAT} and \HESS~are very sensitive to gamma-ray lines from the quintuplet MDM annihilations in the Galactic Center, due to the Sommerfeld-enhanced cross section and the clean, peaked signal~\cite{Cirelli:2015bda}. This candidate may be already ruled out or in the reach of near-future experiments, depending on the DM density in the Galactic Center. We therefore deemed it timely to perform a critical review of the MDM setup, pointing out yet unexplored generalizations of this framework. We propose and explore two distinct directions, discussed in the following. One is to lower the cutoff of the model to allow for DM decays. Another possibility is to abandon the assumption of DM electric neutrality in favor of absolutely stable, millicharged DM candidates.

\section{Decaying quintuplet MDM}

\begin{figure}[t]
\centering
\hspace*{\fill}
\includegraphics[width=0.45\textwidth]{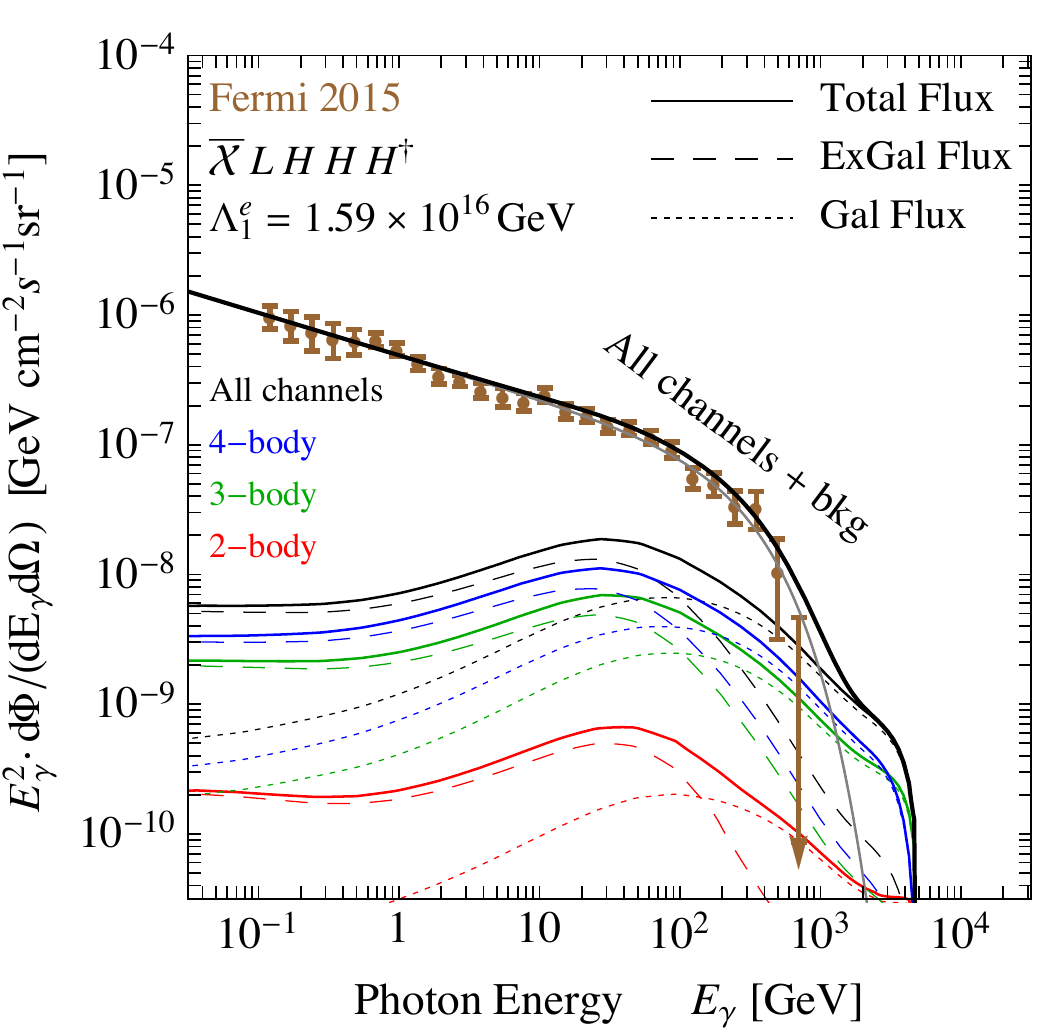}
\hfill
\includegraphics[width=0.45\textwidth]{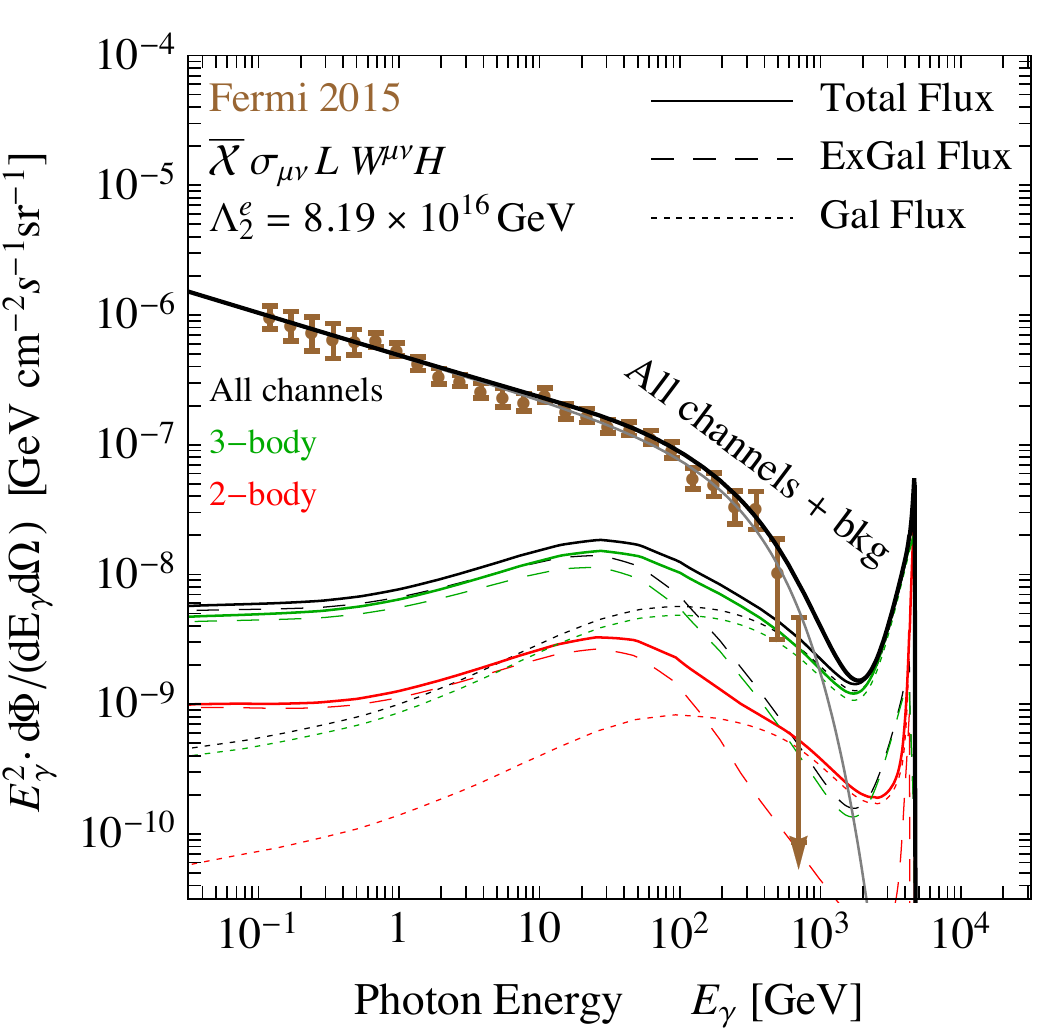}
\hspace*{\fill}
\caption{\label{fig:quintuplet e}\em
Isotropic gamma-ray flux due to DM decays induced by the operators $\Op_1$ (\textbf{left}) and $\Op_2$ (\textbf{right}), assuming DM coupling to electrons and electron neutrinos ($a = e$). \FERMI~data on the diffuse isotropic gamma-ray flux~\cite{Ackermann:2014usa} are shown in brown, and the astrophysical background is displayed as a gray line.}
\end{figure}

The lowest-order Lagrangian operators responsible for breaking the accidental symmetry stabilizing the MDM quintuplet are the two dimension-$6$ operators
\begin{align}
\Op_1 \equiv \frac{c_1^a}{\Lambda^2} \overline{\X} L^a H H H^\dagger \ ,
&&
\Op_2 \equiv \frac{c_2^a}{\Lambda^2} \overline{\X} \sigma_{\mu \nu} L^a W^{\mu \nu} H \ ,
\end{align}
with $H$ the Higgs doublet, $L^a$ the left-handed lepton doublet of flavor $a = e, \mu, \tau$, and $W^{\mu \nu}$ the $\SUd$ gauge boson field strength. For a low enough cutoff $\Lambda$ we should be able to see the products of DM decay. We compute the photon flux from DM decays with the code described in Ref.~\cite{Cirelli:2010xx}, and constrain the cutoff using \FERMI~data on the diffuse isotropic flux~\cite{Ackermann:2014usa} and \HESS~data on gamma-ray lines~\cite{Abramowski:2013ax}. The two operators present a peculiar phenomenology, in that decays into a larger number of particles are favored over final states with fewer particles. In fact, out of each $H$ field, one could either take the Higgs vev $v$, or a Higgs or longitudinal gauge boson. In the first case, one gets one less particle in the final state (and thus a larger phase space), but also a suppression by a factor $(v / \mDM)^2 \approx 10^{-3}$ in the decay rate. Also of interest, $\Op_2$ induces a gamma-ray line-like feature in the spectrum due to $\X \to \gamma \nu$ and other decays with a nearly monochromatic photon. \Fig{fig:quintuplet e} shows the spectral photon flux, broken down in contributions, for the two operators separately. Our take-home messages are:
\begin{itemize}
\item In both cases, the cutoff of the model is constrained to lie above the GUT scale.
\item For $\Op_2$, the best bound is set by the \FERMI~data on the diffuse isotropic flux~\cite{Ackermann:2014usa}, instead of the \HESS~data on gamma-ray lines~\cite{Abramowski:2013ax} as one may have expected.
\item Considering $\Op_1 + \Op_2$, and taking $\Op_2$ to come from a loop-suppressed process as its Lorentz structure may seem to suggest, the gamma-ray line-like feature is dwarfed by the continuum photon flux due to $\Op_1$. Therefore, in general one should consider, beside operators generating gamma-ray lines, also the diffuse emission from other operators arising at the same order in the EFT expansion.
\end{itemize}

\section{Millicharged MDM}
\begin{figure}[t]
\centering
\hspace*{\fill}
\includegraphics[width=0.45\textwidth]{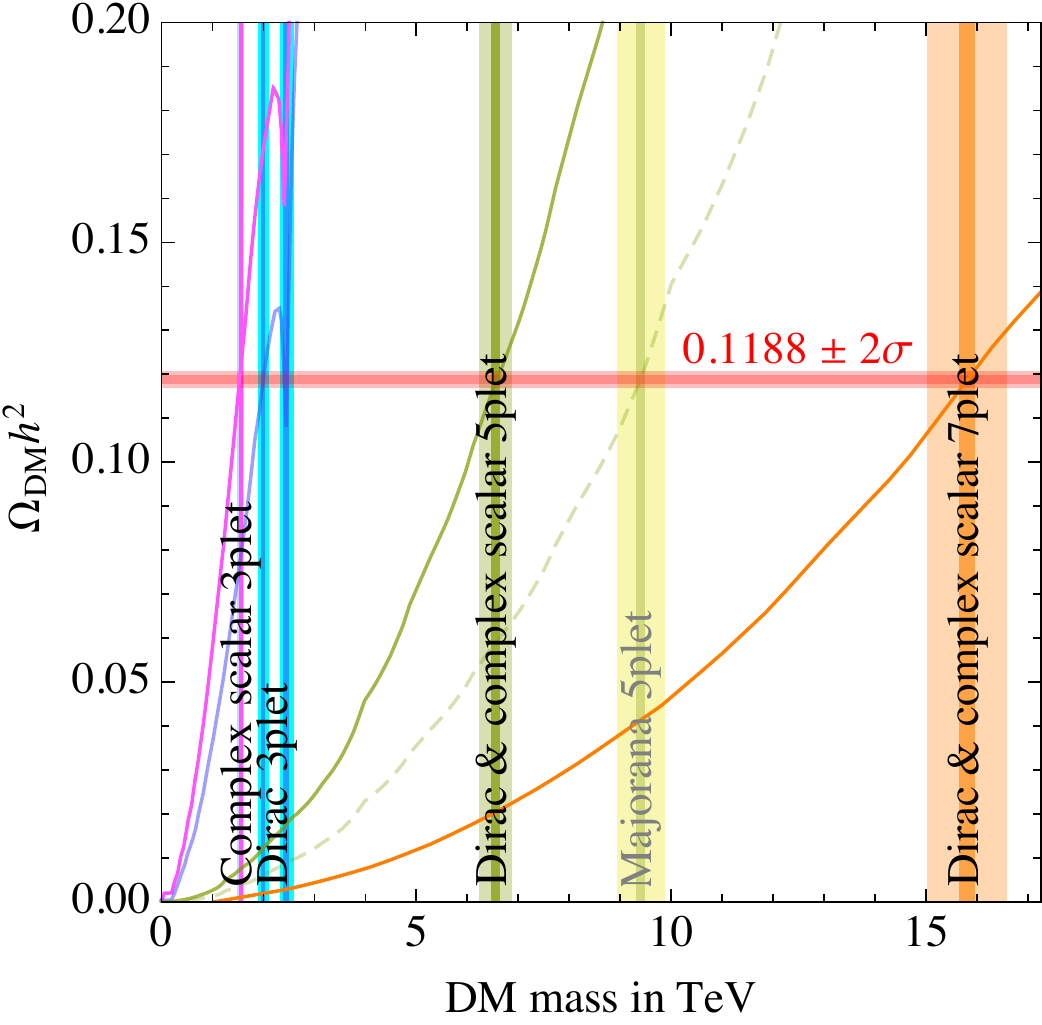}
\hfill
\includegraphics[width=0.45\textwidth]{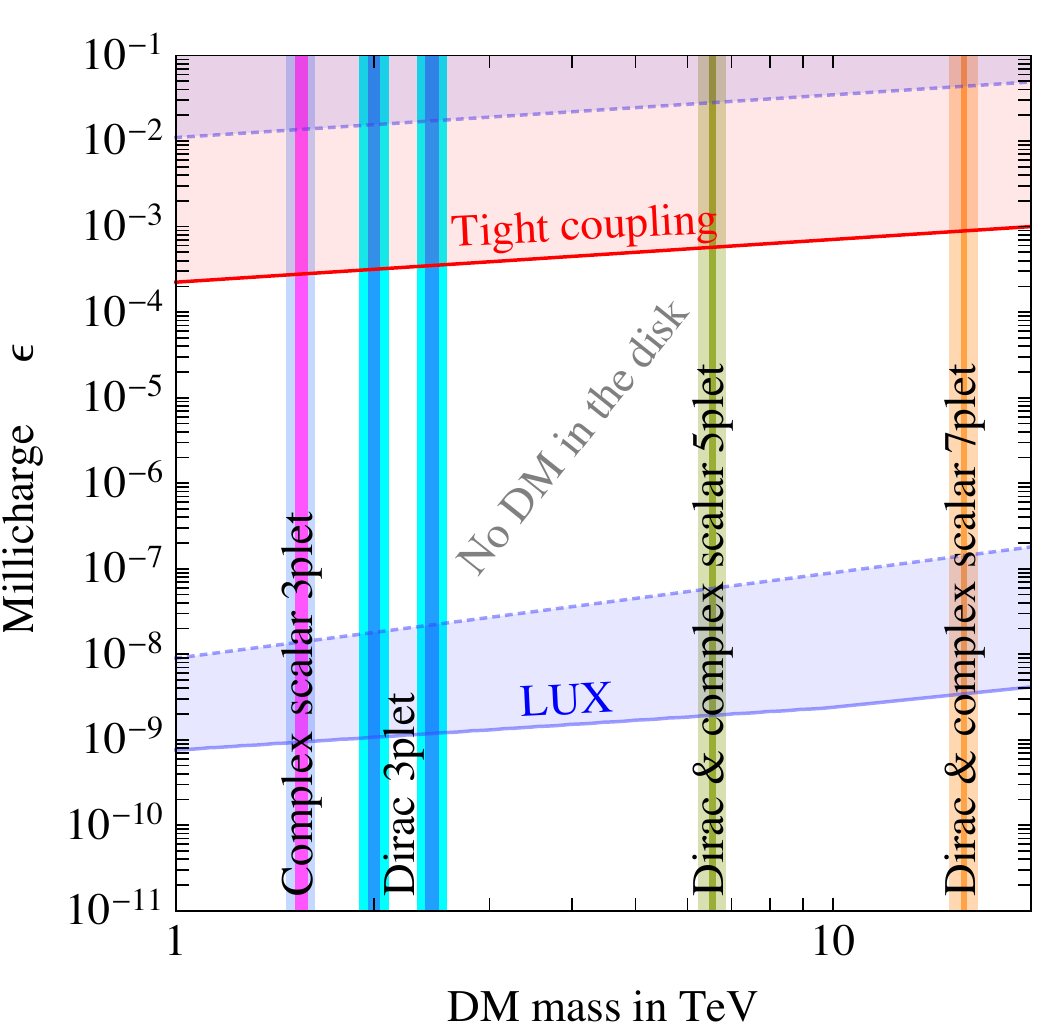}
\hspace*{\fill}
\caption{\label{fig:Masses}\em
\textbf{Left:} Thermal relic abundance and mass of millicharged candidates (the Majorana quintuplet is shown for reference). The relic density line for the Dirac triplet crosses the red band (indicating the measured DM abundance from Ref.~\cite{Ade:2015xua}) twice, thus there are two allowed values for its mass.
\textbf{Right:} Constraints on the absolute value of the DM millicharge as a function of the DM mass.}
\end{figure}

As noted above, the assumption that the DM is electrically neutral singles out a Majorana quintuplet as the only viable MDM candidate. If we give up this assumption and allow for non-zero hypercharge assignments for the new generic multiplet, we get a host of new possible DM candidates. Again we may list all possible multiplets we can add to the SM, but now without the need to worry about DM stability being spoiled by higher-order operators or by Landau poles. In fact, for small (but non-zero) hypercharges, $|\epsilon| \ll 1$ as required by experiments (see \Fig{fig:Masses}), the so-called millicharged DM is made absolutely stable (\ie to all orders in the EFT expansion) by electric charge conservation. So in principle also large-dimensional $\SUd$ representations become viable.

Another welcome feature of a non-zero hypercharge is the following. For odd-dimensional $\SUd$ representations (as required to evade direct DM detection constraints), $\epsilon = 0$ implies that the multiplet is in a real representation of the gauge group, while for $\epsilon \neq 0$ the representation is complex. In the first case the DM is self-conjugated while in the second case it is not. Therefore, a non-zero hypercharge implies a doubling of degrees of freedom, which changes the relic density (and therefore the DM mass) as well as the bounds from indirect searches. Under some conditions~\cite{DelNobile:2015bqo}, we can now describe the DM as composed of two mutually decoupled species with same mass and interactions, and therefore the relic density for $\epsilon \neq 0$ is twice that for $\epsilon = 0$. Moreover, since now particles and anti-particles are distinct, the probability a DM particle finds a partner to annihilate with is half that for a self-conjugated DM, thus making indirect detection bounds less stringent for a non self-conjugated candidate.

The left panel of \Fig{fig:Masses} shows the relic density and mass of few of our millicharged DM candidates (for reference, the Majorana quintuplet is the standard, neutral MDM candidate). The right panel instead displays bounds on the absolute value of the DM millicharge from CMB observations~\cite{Dolgov:2013una} and \LUX~\cite{Akerib:2013tjd, DelNobile:2013sia, Chuzhoy:2008zy}. When faced with gamma-ray line searches in the Galactic Center, most of our millicharged candidates perform as good (or bad) as the MDM Majorana quintuplet: they are excluded (allowed) for a cuspy (cored) DM profile. However, we find that a Dirac triplet with mass about $2$ TeV is a viable candidate even for a cuspy profile.

\section{Conclusions}
Minimal Dark Matter~\cite{Cirelli:2005uq, Cirelli:2007xd, Cirelli:2009uv} is an elegant and extremely predictive framework, which singles out the neutral component of a spin-$\frac{1}{2}$ $\SUd$ quintuplet with zero hypercharge as the DM. Present day gamma-ray line searches in the Galactic Center are particularly sensitive to this candidate, thus making it timely to perform a critical review of the model to find possible yet unexplored generalizations.

We proposed and explored two distinct directions~\cite{DelNobile:2015bqo}. One is to lower the cutoff of the model to allow for decays of the DM quintuplet. A careful analysis of the decay spectrum of this candidate showed that current gamma-ray data constrain the cutoff to lie above the GUT scale. Another possibility is to abandon the assumption of DM electric neutrality in favor of absolutely stable, millicharged DM candidates. We found that a Dirac $\SUd$ triplet with mass around $2$ TeV is a viable candidate still unconstrained by the stringent bounds from gamma-ray line searches.

\end{document}